\documentclass[12pt]{iopart}

\usepackage{iopams}
\usepackage{graphicx}
\usepackage{latexsym}

\newtheorem{theorem}{Theorem}

\usepackage{color}

\begin{document}

\title[Preserving Information from \ldots]{Preserving information from the beginning to the end of time in a
Robertson-Walker spacetime}

\author{Stefano Mancini$^{1,2}$, Roberto Pierini$^{1,2}$ and Mark M. Wilde$^3$}

\address{$^1$School of Science and Technology, University of Camerino, 62032 Camerino, Italy\\
$^2$INFN Sezione di Perugia, 06123 Perugia, Italy\\
$^3$Hearne Institute for Theoretical Physics, Department of Physics and Astronomy,
Center for Computation and Technology,
Louisiana State University, Baton Rouge, Louisiana 70803, USA}
\ead{stefano.mancini@unicam.it}

\begin{abstract}
Preserving information stored in a physical system subjected to noise can be modeled
in a communication-theoretic paradigm, in which
storage and retrieval correspond to an input
encoding and output decoding, respectively. The encoding and decoding are then constructed
in such a way as to protect against the action
of a given noisy quantum channel. This paper considers the situation
in which the noise is not due to technological imperfections, but rather to the
physical laws governing the evolution of the universe. In particular, we
consider the dynamics of quantum systems under a 1+1 Robertson-Walker spacetime and
find that the noise imparted to them is equivalent to the well known amplitude
damping channel. Since one might be interested in preserving both classical
and quantum information in such a scenario, we study trade-off 
 coding strategies and determine a region of
achievable rates for the preservation of both kinds of information.
For applications beyond the physical setting studied here, we also determine a trade-off between
achievable rates of classical and quantum information preservation
when entanglement assistance
is available.
\end{abstract}

\vspace{2pc}

\submitto{\NJP}

\section{Introduction}

Data storage is relevant not only for accomplishing tasks in our day-to-day
lives but also to keep track of our history. Information can be stored on
various physical media, ranging from modern compact disks to ancient papyrus.
A fundamental goal of information storage is to preserve it for the longest
possible time in a reliable way. That obviously depends on the used
technology. However, even if we have achieved a perfect or ideal implementation of a given
technology, we should realize that there are limitations on preserving information.
These limitations are posed by
physical theories and ultimately result from the evolution of the universe itself,
which can cause unavoidable effects to any physical system.

To address the issue of determining these fundamental limits, we require the
theory of general relativity as well as that of quantum information. Very
recently the interconnections between these two fields have received increased
interest. Several previous works have developed a theory of communication
between a sender and a receiver in relativistic settings \cite{BHP09, BHTW10, JBW,  MHM, BOJ,
BHP} or in situations involving black holes \cite{BA, HP}.

In this paper, we investigate how well
information stored in the remote past is preserved when going to the far
future, by assuming evolution of the universe in a Robertson-Walker (RW) spacetime.
Our main results are 1) that the noise imparted to spin-$\frac{1}{2}$ particles
by the evolution of the universe
is equivalent to an amplitude damping channel, and so we then 2) determine achievable rates
for the simultaneous communication of classical and quantum information over this channel.
As a result, we can interpret these rates to be achievable rates for the storage
of classical and quantum information from the early past to the far future in a Robertson-Walker spacetime.

The RW spacetimes are a reasonable description of the dynamics of
the late universe, which, at large scale, appear to be homogeneous and
isotropic. Most cosmological models are special cases of RW
spacetimes \cite{BD}. When considering a quantum matter field evolving through
a dynamical spacetime, the concept of the vacuum cannot be considered unique any
longer. Indeed, to detect the presence of quanta it is also necessary to
specify the details of the quantum measurement process, and in particular, the
state of motion of the measuring device. Particles possess an essential
observer-dependent quality, so that they can be observed on some detectors and
not others. Also, we can define positive and negative energy solutions of
differential equations governing matter fields only if the spacetime structure
is invariant under the action of a time-like Killing vector field \cite{WAb}.
This is certainly true for Minkowski spacetime, and a RW universe which is
Minkowskian in the early past and in the far future is a suitable choice.
The simplest, nevertheless insightful, choice we can make is a 1+1 RW spacetime.

There, we can consider any quantum state of the matter field before the expansion of
the universe begins and define, without ambiguity, its particle content. We
then let the universe expand and check how the state looks once the expansion
is over. The overall picture can be thought of as a noisy channel into which
some quantum state is fed. Once we have defined the quantum channel emerging from the
physical model, we will be looking at the usual communication task as
information transmission over the channel.
Since we are interested in the preservation of any kind of information, we
shall consider the trade-off between the different resources of classical and
quantum information.

The rest of the paper is organized as follows. In the next section, we discuss the physical model
and show how the noise imparted to spin-$\frac{1}{2}$ particles is equivalent to
an amplitude damping channel, which has been well studied in quantum
information theory \cite{W13book}. In the section thereafter, we calculate achievable rates
for the simultaneous communication of classical and quantum information over this channel.
We then conclude with a summary of our results and a discussion of some open questions.  
Appendixes A and B are devoted to prove the main results.  
There we also determine a trade-off between
achievable rates of classical and quantum information preservation
when entanglement assistance is available, which might be useful for 
applications beyond the physical setting studied in this paper.


\section{The physical model}

\subsection{Robertson-Walker spacetime}

The geometry of a RW spacetime, considered here for the sake of simplicity of 1+1 dimensions,
is described by the line element
\begin{equation*}
\label{line}ds^{2}=[a(\tau)]^{2}(-d\tau^{2}+dx^{2}),
\end{equation*}
which represents the varying distance between two spacetime points, depending
on the conformal scale factor $a(\tau)$. The so-called conformal time $\tau$
is a function depending on the cosmological time~$t$, defined by $\tau=\int
a^{-1}(t)dt$. The spatial coordinate is denoted by $x$. Now consider an
expanding universe with Minkowskian spacetime in the early past and in the far
future filled with a Dirac field (that is, with matter made of spin-$\frac{1}{2}$
particles). We can associate a Hilbert space to each of the two regions with
suitable basis vectors. The dynamics of the matter fields $\psi$ of mass $m$
are governed by the Dirac equation expressed in covariant form
\begin{equation}
\label{Deq}(\tilde\gamma^{\mu}D_{\mu}+m)\psi=0.
\end{equation}
The index $\mu$ runs from $0$ to $1$ and 
the Einstein sum rule over repeated indices is used. Furthermore, 
$\tilde\gamma^{\mu}\equiv[a(\tau)]^{-1}\gamma^\mu$ with $\gamma^\mu$ the $2\times 2$ matrices representing Dirac algebra.
Finally, $D_{\mu}$ is the covariant derivative \cite{BD}.

We look for solutions of (\ref{Deq}), writing $\psi=a^{-1/2}(\gamma^{\nu
}\partial_{\nu}-M)\varphi$, with $M=ma(\tau)$, so to have
\begin{equation}
g^{\mu\nu}\partial_{\mu}\partial_{\nu}\varphi-\gamma^{0}\dot{M}\varphi
-M^{2}\varphi=0, \label{newDeq}%
\end{equation}
with $g^{\mu\nu}$ being the flat metric as opposed to the actual spacetime
metric $\tilde{g}^{\mu\nu}=[a(\tau)]^{-2}g^{\mu\nu}$. 
Moreover, given flat spinors 
$u$ and $v$ satisfying $\gamma^0u=-iu$ and $\gamma^0v=iv$, we set
\begin{eqnarray}
&  \varphi^{(-)}\equiv N^{(-)}f^{(-)}(\tau) u e^{ i k x}, \label{varphi1}\\
&  \varphi^{(+)}\equiv N^{(+)}f^{(+)}(\tau) v e^{ i k x}, \label{varphi2}%
\end{eqnarray}
with $k$ the momentum. Inserting (\ref{varphi1}), (\ref{varphi2}) into (\ref{newDeq}), the functions $f^{(\pm)}$ 
must obey the differential equation
\begin{equation}
\ddot{f}^{(\pm)}+\left(  k^{2}+M^{2}\pm i\dot{M}\right)
f^{(\pm)}=0. \label{feq}%
\end{equation}
Define  $f^{(\pm)}_{in/out}$ and ${f^{(\pm)}}^*_{in/out}$ the solutions behaving 
as positive and negative frequency modes
with respect to conformal time~$\tau$ near the asymptotic past/future,
i.e. $\dot{f}^{(\pm)}_{in/out}(\tau)\approx - iE_{in/out}f^{(\pm)}_{in/out}(\tau)$ with
\begin{equation*}
E_{in/out}\equiv\sqrt{k^{2}+M_{in/out}^{2}},\quad M_{in/out}\equiv m a(\tau\to+/-\infty). \label{Einout}%
\end{equation*}
Then we can introduce spinors that behave like positive and
negative energy spinors, respectively, in the asymptotic regions:
\begin{eqnarray}
U_{in/out}(k,x,\tau) &  \equiv
N^{(-)}(\gamma^{\nu}\partial_{\nu}-M) f^{(-)}_{in/out}(\tau) u e^{ikx},\label{Updeta}\\
V_{in/out}(k,x,\tau)  &  \equiv 
 N^{(+)} (\gamma^{\nu}\partial_{\nu}-M) {f^{(+)}}^*_{in/out}(\tau) v e^{-ikx}, \label{Vpdeta}%
\end{eqnarray}
with normalization constants
\begin{equation*}
N^{(+)}=N^{(-)}=\left[\frac{E_{in/out}-M_{in/out}}{2k^2M_{in/out}}\right]^{1/2}.
\end{equation*}
Now the solutions of (\ref{Deq}) can be expanded over (\ref{Updeta}) and (\ref{Vpdeta}) as
\begin{eqnarray}
\psi(x,\tau)  &  = \int dk \; a^{-1/2}(\tau)\left[a_{in}(k)U_{in}(k,x,\tau)
+b_{in}^{\dag}(k)V_{in}(k,x,\tau)\right],\quad\mathrm{or}\nonumber\\
\psi(x,\tau)  &  =\int dk\; a^{-1/2}(\tau)\left[ a_{out}(k)U_{out}(k,x,\tau)
+b_{out}^{\dag}(k)V_{out}(k,x,\tau)\right].\nonumber
\end{eqnarray}
The coefficients appearing in such expansions are $in$-$out$ ladder operators 
for particles and antiparticles ($a$, $a^\dag$ and $b$, $b^\dag$ respectively).
They are connected by Bogoliubov transformations \cite{DUN}
\begin{eqnarray}
a_{out}( k)&=\alpha(k)\,a_{in}( k)-\beta( k)\,b_{in}^{\dag}(-k),
\label{B1}\\
b_{out}^{\dag}(-k)&=\beta^\ast(k)\,a_{in}(k)+\alpha^\ast(k)\,b_{in}^{\dag}(-k),
\label{B2}
\end{eqnarray}
where $\alpha$, $\beta\in\mathbb{C}$, such that
$|\alpha|^{2}+|\beta|^{2}=1$ and $\alpha\beta^{\ast}-\alpha^{\ast}\beta=0$.
Notice that such transformations do
not mix different momentum solutions, and so we can safely focus on a single momentum
and omit the dependence on $k$. Therefore, any particle (antiparticle) 
quantum state lives in a 2-dimensional Hilbert space with orthonormal basis $\{ |0\rangle,
|1\rangle\}$ denoting absence or presence of a particle (antiparticle). 
The Bogoliubov coefficients are linked
to physical quantities by $|\beta|^{2}=n/2$, where $n$ is the density of particles for the
mode under consideration ($0\le n\le2$).

The transformations (\ref{B1}), (\ref{B2}) come
from a unitary operator that can be written as
\begin{equation}
U=\exp\left[  r\left(  e^{-i\vartheta}b^{\dag}_{in}a_{in}^{\dag}-e^{i\vartheta}
a_{in}b_{in}\right)  \right] ,
\label{Unitary}
\end{equation}
where the parameters $r$ and $\vartheta$ are related to $\alpha$ and $\beta$ of Eqs.(\ref{B1}) 
and (\ref{B2}) by  $\alpha=\cos r$ and $\beta=-e^{-i\vartheta}\sin r$.

\subsection{Robertson-Walker dynamics induces amplitude damping channel}

Assuming to have access to particles in the out region only, the antiparticles
will play the role of an environment which is initially
in the vacuum. Hence, from (\ref{Unitary}), we
can single out a completely positive trace preserving linear
map from $in$ particle states to $out$ particle states, given
by
\begin{equation}
\rho\mapsto\mathcal{A}(\rho)=\mathrm{tr}_{-p}\left[  U\left(  \rho
\otimes|0\rangle_{-p}\langle0|\right)  U^{\dag}\right]  \,, \label{kraus-rep}%
\end{equation}
where $U$ is given by (\ref{Unitary}) and $\mathrm{tr}_{-p}$ stands for the
partial trace over antiparticles. In terms of the so-called Kraus representation,
we have that
\begin{equation*}
\mathcal{A}(\rho)=\sum_{j=0,1}K_{j}\rho K_{j}^{\dag},
\end{equation*}
where $K_{j}={}_{-p}\langle j|U|0\rangle_{-p}$ follows from (\ref{Unitary})
\begin{eqnarray}
K_{0}  &  =I+(\cos r-1)a_{in}a_{in}^{\dag},\nonumber\\
K_{1}  &  =e^{-i\vartheta} \sin r  \,a_{in}^{\dag}.\nonumber
\end{eqnarray}
Expressing them in terms of outer products of particles states
\begin{eqnarray}
K_{0}  &  = \left\vert 1\right\rangle \left\langle 1\right\vert
+\sqrt{\eta}\left\vert 0\right\rangle \left\langle 0\right\vert, \\
K_{1} & = \sqrt{1-\eta}\left\vert 1\right\rangle
\left\langle 0\right\vert ,
\label{Kraus-A}%
\end{eqnarray}
we may observe that the quantum channel map
$\mathcal{A}$ is an amplitude damping channel with
the so-called \emph{transmissivity} $\eta\in[0,1]$ related to the
physical observable $n$ by
\begin{equation}
\eta=\cos^{2}r=1-\frac{n}{2}. \label{eta}
\end{equation}


We now consider the toy model introduced in \cite{DUN}, having the following
conformal scale factor
\begin{equation}
\label{scalefactor}a(\tau)=1+\epsilon(1+\tanh\rho\tau) \,.
\end{equation}
The real and positive parameters $\epsilon$ and $\rho$ control the total
volume and the rapidity of expansion of the universe, respectively.
In the two asymptotic regions $in$ and
$out$, we have that $a(\tau\rightarrow-\infty)=1$ and $a(\tau\rightarrow
+\infty)=1+2\epsilon$, respectively.

Inserting (\ref{scalefactor}) into (\ref{feq}) we get
\begin{equation}
\ddot{f}^{(\pm)}+\left[  k^{2}+m^{2}(1+\epsilon(1+\tanh\rho
\tau))^{2}\pm\frac{im\rho\epsilon}{\cosh^{2}\rho\tau}\right]  f^{(\pm
)}=0.\nonumber
\end{equation}
Solutions of this equation can be found as \cite{DUN}
\begin{eqnarray}
&  f_{in}^{(\pm)}(\tau)=e^{(-iE_{+}\tau-\frac{i}{\rho}E_{-}\ln(2\cosh\rho
\tau))}\nonumber\\
&\hspace{1.3cm}\times\ _{2}{\sf F}_{1}\left(  1+i\frac{E_{-}\pm m\epsilon}{\rho},i\frac{E_{-}\mp
m\epsilon}{\rho},1-i\frac{E_{in}}{\rho},\frac{1+\tanh\rho\tau}{2}\right)  ,\nonumber\\
&  f_{out}^{(\pm)}(\tau)=e^{(-iE_{+}\tau-\frac{i}{\rho}E_{-}\ln(2\cosh\rho
\tau))}\nonumber\\
&\hspace{1.3cm}\times\ _{2}{\sf F}_{1}\left(  1+i\frac{E_{-}\pm m\epsilon}{\rho},i\frac{E_{-}\mp
m\epsilon}{\rho},1+\frac{i}{\rho}E_{out},\frac{1-\tanh\rho\tau}{2}\right)  \,,\nonumber
\end{eqnarray}
with ${}_{2}{\sf F}_{1}$ denoting the ordinary hypergeometric function and
\begin{equation*}
E_{\pm}\equiv\frac{E_{out}\pm E_{in}}{2}\,.
\end{equation*}

Since $f^{(\pm)}_{in/out}(\tau)$ and ${f^{(\pm)}}_{in/out}^*(\tau)$ are positive and negative frequency
modes in asymptotic regions, we can write the Bogoliubov transformation
between them as follows:
\begin{equation*}
f_{in}^{(\pm)}(\tau)=A^{(\pm)}(k)f_{out}^{(\pm)}(\tau)+B^{(\pm
)}(k){f_{out}^{(\mp)}}^*(\tau)\,.
\end{equation*}
Using linear transformation properties of hypergeometric functions we can
write down the coefficients as \cite{DUN}
\begin{eqnarray}
A^{(\pm)}(k)  &  =\frac{\Gamma(1-\frac{i}{\rho}E_{in})\Gamma
(-\frac{i}{\rho}E_{out})}{\Gamma(1-\frac{i}{\rho}E_{+}\pm\frac{im\epsilon
}{\rho})\Gamma(-\frac{i}{\rho}E_{+}\mp\frac{im\epsilon}{\rho})},\nonumber\\
B^{(\pm)}(k)  &  =\frac{\Gamma(1-\frac{i}{\rho}E_{in})\Gamma(\frac
{i}{\rho}E_{out})}{\Gamma(1+\frac{i}{\rho}E_{-}\pm\frac{im\epsilon}{\rho
})\Gamma(\frac{i}{\rho}E_{-}\mp\frac{im\epsilon}{\rho})},\nonumber
\end{eqnarray}
with $\Gamma$ denoting the Euler Gamma function.
These Bogoliubov coefficients will be related to those of Eqs.(\ref{B1}) and (\ref{B2}), 
namely $\alpha$ and $\beta$~\cite{MPM14}.
In particular it results
\begin{equation*}
|\alpha(k)|^{2}=\frac{E_{out}(E_{in}-M_{in})}{E_{in}(E_{out}-M_{out})}
\,|A^{(-)}(k)|^{2},
\end{equation*}
Hence, remembering from (\ref{eta}) that $\eta=1-\frac{n}{2}=|\alpha|^{2}$,
we find 
\begin{eqnarray}
\eta=\frac{E_{out}(E_{in}-M_{in})}{E_{in}(E_{out}-M_{out})}
\,\Bigg| \frac{\Gamma(1-\frac{i}{\rho}E_{in})\Gamma
(-\frac{i}{\rho}E_{out})}{\Gamma(1-\frac{i}{\rho}E_{+}-\frac{im\epsilon
}{\rho})\Gamma(-\frac{i}{\rho}E_{+}+\frac{im\epsilon}{\rho})} \Bigg|^{2} \,.
\label{transm}
\end{eqnarray}

\begin{figure}[ptb]
\centering
\includegraphics[width=4.0in]{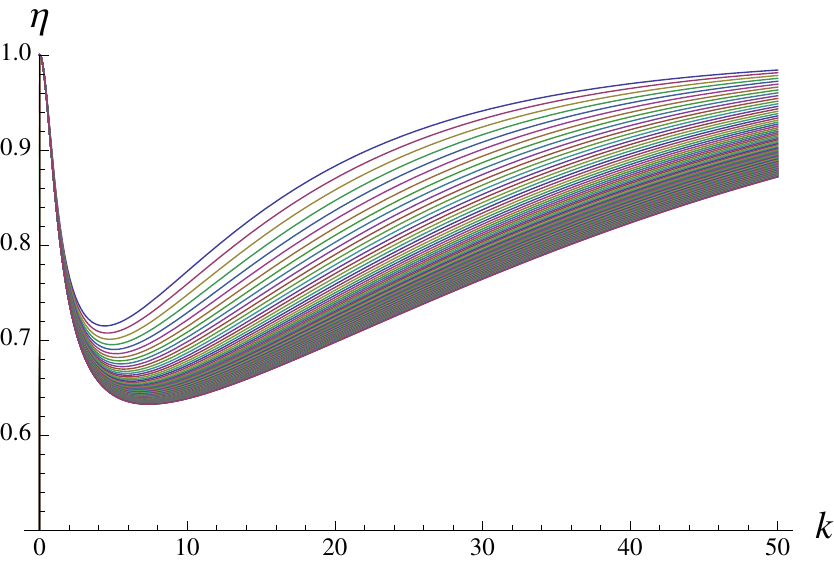}\caption{Transmissivity $\eta$ vs.
momentum $k$ for $\epsilon$ ranging from $10$ (top curve) to $100$ (bottom curve) by step $1$.
The values of other parameters are $\rho=100$ and $m=1$.}%
\label{fig:damping}%
\end{figure}

In Figure~\ref{fig:damping}, we plot the transmissivity $\eta$ in (\ref{transm}) as
a function of the momentum $k$. Observe that it is equal to one
(no damping) only for zero or large momentum. This is a consequence of the
fact that modes such that $0<\sqrt{k^{2}+m^{2}}<\rho$ are excited, implying
particle creation for them.
Also notice that the value of $\eta$ never drops below $1/2$,
and it is equal to this minimum value in the limit as $\rho,\epsilon\to\infty$.

\section{Information trade-offs for the amplitude damping channel}

In the above development, observe that the region for which $\eta$ falls below
one is the most important for information storage. In fact, in order to save
energy, one would like to have the momentum $k$ as low as possible. However, it is
unreasonable to freeze particles such that $k=0$. Hence, we have to face up
with the problem of non-negligible information damping, and this motivates us
to consider the best strategy for preserving it.

In particular, we would like
to preserve both classical and quantum information in the RW spacetime, and so
we consider trade-off strategies for doing so \cite{cmp2005dev,WH10b},
modeling the noise as an amplitude
damping channel (as motivated in the previous section).
To do so, we can model this
problem in a communication-theoretic language, in which we say that the device
encoding information at the beginning of the evolution is the ``sender'' and
the device recovering information at the end of the evolution is the ``receiver.''

A simple strategy for trading between classical and quantum communication
is known as {\it time sharing}---in a time-sharing strategy, the sender and receiver
use a classical communication code for a fraction of
the channel uses, a quantum communication code for another fraction, etc. 
For some channels such as
the quantum erasure channel \cite{GBP97},
time sharing is an optimal communication strategy, but in general,
it cannot outperform a more general strategy known as
``trade-off coding''  \cite{WH10b}.
This allows for transmitting classical and quantum
information at net rates $(C,Q)$ that lie in a two-dimensional capacity region.

To proceed with our development for the amplitude damping channel,
we begin by recalling that the trade-off region between classical and
quantum communication (without the help of entanglement assistance) 
for any quantum channel 
$\mathcal{A}_{A^{\prime}\rightarrow B}$\ is given by  \cite{cmp2005dev}:
\begin{eqnarray}
Q  &  \leq I\left(  A\rangle BX\right)  _{\rho},
\label{eq:double-1}\\
C+Q  &  \leq I\left(  X;B\right)  _{\rho}+I\left(  A\rangle BX\right)_{\rho},
\label{eq:double-2}
\end{eqnarray}
where $I(AX;B)_\rho \equiv H(AX)_\rho + H(B)_\rho - H(ABX)_\rho$,
$I(A\rangle BX)\equiv H(BX)_\rho - H(ABX)_\rho $, and
$I(X;B)_\rho\equiv H(X)_\rho + H(B)_\rho - H(BX)_\rho$ denote the quantum
mutual information, coherent information, and Holevo information of a quantum state $\rho_{XAB}$, 
respectively, with the von Neumann entropies defined
as $H(A)_\rho \equiv -\rm{Tr} \{ \rho_A \log \rho_A\}$,
 $H(B)_\rho \equiv -\rm{Tr} \{ \rho_B \log \rho_B\}$,
 $H(AB)_\rho \equiv -\rm{Tr} \{ \rho_{AB} \log \rho_{AB}\}$, etc.~(see Chapter 11 of \cite{W13book},
for example, for more on these definitions).
These entropies are actually with
respect to a classical-quantum state of the following form:
\begin{equation}
\rho_{XAB}\equiv\sum_{x}p_{X}\left(  x\right)  \left\vert x\right\rangle
\left\langle x\right\vert _{X}\otimes\mathcal{A}_{A^{\prime}\rightarrow
B}\left(  |\phi^{x}\rangle_{AA^{\prime}}\langle\phi^{x}|\right)  ,
\label{rhoXAB}%
\end{equation}
with $|\phi^{x}\rangle_{AA^{\prime}}\langle\phi^{x}|$ a purification of the
input state $\rho_{A^{\prime}}^{x}$ corresponding to the letter $x$.
Taking the union of the region specified by (\ref{eq:triple-1}%
)-(\ref{eq:triple-3}) over all ensembles of the form $\left\{  p_{X}\left(
x\right)  , |\phi^{x}\rangle_{AA^{\prime}}\langle\phi^{x}| \right\}  $ then
gives what is known as the single-letter triple trade-off region (meaning that the formulas
are a function of a single instance of the channel). 
We should clarify that the above rate region is an achievable rate region,
and for some channels, it
is known to be optimal as well \cite{BHTW10,WH10b}. The above rate region is not known to be
optimal for the amplitude damping channel. 

For the amplitude damping channel $\mathcal{A}$, and hence for the channel
(\ref{kraus-rep}), we have the following characterization of the
single-letter trade-off region:


\begin{theorem}
\label{theo1} 
The single-letter trade-off region  (\ref{eq:double-1})-(\ref{eq:double-2}) for the qubit amplitude damping channel 
is the union of the following polyhedra over all $p_0, q_x,\nu_x\in\left[  0,1\right]$ for $x \in \{0,1\}$ and where 
$p_1 = 1 - p_0$ and
$p \equiv \sum_{x\in\{0,1\}} p_x q_x$:
\begin{eqnarray}
Q &  \leq  \sum_{x\in\{0,1\}} p_x \, [{\sf g}\left(  q_x,\eta,\nu_x\right)  -{\sf g}\left(  q_x,1-\eta,\nu_x\right)],
\nonumber\\
C+Q &  \leq  h_{2}\left(  \eta p\right)  -\sum_{x\in\{0,1\}} p_x  \, {\sf g}\left(  q_x,1-\eta,\nu_x\right).
\nonumber
\end{eqnarray}
Furthermore
\begin{equation*}
{\sf g}\left(  q,z,\nu\right)  \equiv  h_{2}\left(  \frac{1+\sqrt{\left(
1-2 z q\right)^{2}+4 z \nu^2 q (1-q)  }}{2}\right),
\end{equation*}
with $h_{2}$ denoting the binary Shannon entropy:  $h_{2}(y)\equiv  -y\log_2 y-(1-y)\log_2(1-y)$, $y\in[0,1]$.
\end{theorem}

The proof of Theorem~\ref{theo1} is given in \ref{sec:theorem1-proof}.
We can significantly simplify the characterization of the region when $\eta \geq 1/2$,
which is the case of most interest for the physical setting of this paper.

\bigskip

\begin{theorem}
\label{theo2} 
The single-letter trade-off region  (\ref{eq:double-1})-(\ref{eq:double-2}) for the qubit amplitude damping channel 
when $\eta\geq 1/2$ is the union of the following polyhedra over all $p,\nu\in\left[  0,1\right]$:
\begin{eqnarray}
Q &  \leq {\sf g}\left(  p,\eta,\nu\right)  -{\sf g}\left(  p,1-\eta,\nu\right)  ,\nonumber\\
C+Q &  \leq h_{2}\left(  \eta p\right)  -{\sf g}\left(  p,1-\eta,\nu\right) ,\nonumber
\end{eqnarray}
where ${\sf g}\left(  p,z,\nu\right)$ is defined in Theorem~\ref{theo1}
and it can be
achieved with the following ensemble 
\begin{eqnarray}
&\frac{1}{2}\left\vert 0\right\rangle \left\langle 0\right\vert _{X}\otimes%
\left(\begin{array}{cc}
1-p & \nu\sqrt{p\left(  1-p\right)  }\\
\nu\sqrt{p\left(  1-p\right)  } & p
\end{array}\right)
_{A^{\prime}}\nonumber\\
&+\frac{1}{2}\left\vert 1\right\rangle \left\langle 1\right\vert
_{X}\otimes
\left(\begin{array}{cc}
1-p & -\nu\sqrt{p\left(  1-p\right)  }\\
-\nu\sqrt{p\left(  1-p\right)  } & p
\end{array}\right)
_{A^{\prime}},\nonumber
\label{optimal}
\end{eqnarray}
with $p,\nu\in\left[  0,1\right]  $.
\end{theorem}

The proof of Theorem~\ref{theo2} is given in \ref{sec:theorem2-proof}.

\begin{figure}[ptb]
\centering
\includegraphics[width=4.6in]{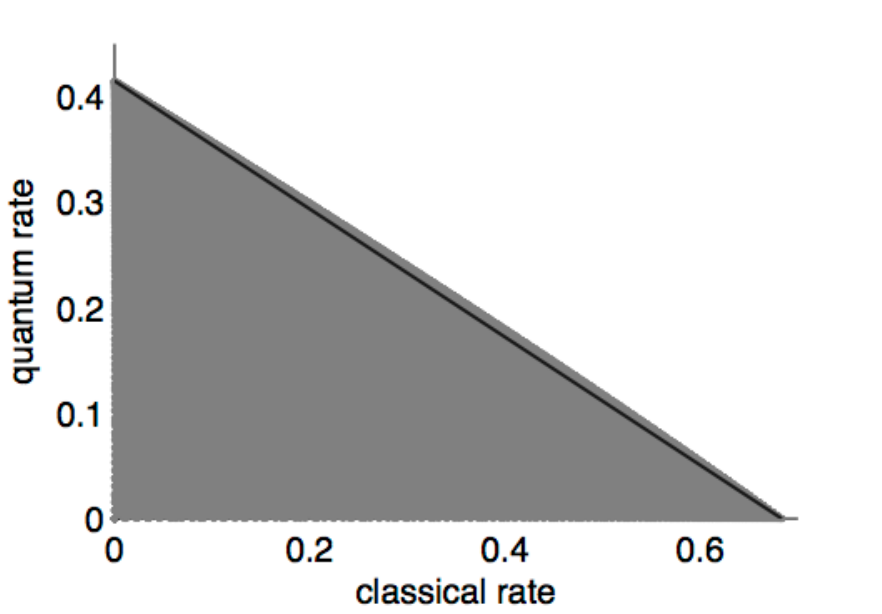}\caption{A comparison of a
trade-off coding strategy (blue points)\ versus a time-sharing strategy (red
line)\ for an amplitude damping channel with transmissivity $\eta=0.75$. The
figure demonstrates that an ensemble of the form in Theorem \ref{theo2}
outperforms a naive time-sharing strategy between the product-state classical
capacity and the quantum capacity.}%
\label{fig:cq-tradeoff}%
\end{figure}

Notice that the ensemble that attains the trade-off interpolates between the strategy that achieves the quantum
capacity of the amplitude damping channel and that which achieves the
product-state classical capacity of the amplitude damping channel, as $\nu$
varies from zero to one. That is, when $\nu=1$, the ensemble reduces to
\begin{eqnarray}
&\frac{1}{2}\left\vert 0\right\rangle \left\langle 0\right\vert _{X}\otimes%
\left(\begin{array}{cc}
1-p & \sqrt{p\left(  1-p\right)  }\\
\sqrt{p\left(  1-p\right)  } & p
\end{array}\right)
_{A^{\prime}}\nonumber\\
&+\frac{1}{2}\left\vert 1\right\rangle \left\langle 1\right\vert
_{X}\otimes
\left(\begin{array}{cc}
1-p & -\sqrt{p\left(  1-p\right)  }\\
-\sqrt{p\left(  1-p\right)  } & p
\end{array}\right)
_{A^{\prime}},\nonumber
\end{eqnarray}
which has been proved to be optimal for the product-state classical capacity
(the single-letter classical capacity)
\cite{GF}. When $\nu=0$, the ensemble reduces to%
\begin{equation*}
\left(  \frac{1}{2}\left\vert 0\right\rangle \left\langle 0\right\vert
_{X}+\frac{1}{2}\left\vert 1\right\rangle \left\langle 1\right\vert
_{X}\right)  \otimes
\left(\begin{array}{cc}
1-p & 0\\
0 & p
\end{array}\right)
_{A^{\prime}},
\end{equation*}
which is of the diagonal form that achieves the quantum capacity of the
amplitude damping channel~\cite{GF}.
The communication strategy resulting from the state in (\ref{optimal}) is very
different from a naive time-sharing one and outperforms it (see
Figure~\ref{fig:cq-tradeoff}).


\section{Discussions and Conclusions}

In this paper, we have investigated how well
information stored in the remote past is preserved when going to the far
future, by assuming evolution of the universe in a Robertson-Walker spacetime.
We proved, under certain assumptions, that the noise imparted to spin-$\frac{1}{2}$ particles
by the evolution of the universe
is equivalent to an amplitude damping channel, and we then determined achievable rates
for the simultaneous communication of classical and quantum information over this channel.
Actually we have established an achievable rate region (and the ensemble to attain it) characterizing communication trade-offs for the qubit amplitude
damping channel, thus also generalizing the results given in
Ref.~\cite{GF}. 
Our results refer to single-letter rate regions, so that it remains open
to determine whether a multi-letter characterization could achieve strictly higher rates
of communication. For this purpose, one might consider recent approaches developed
in \cite{BEHY}.

A more physically relevant scenario is the $3+1$ dimensional spacetime with the same evolutionary model adopted here. In this situation spin degrees of freedom of the quantum field become relevant, making physics somehow more involved but richer. An extension of our study to this case is foreseeable thanks to 
the Bogolyubov transformations given in Ref.\cite{DUN}.
Still we are supposing that the \emph{in} and \emph{out} regions spacetime admits natural particle states and a privileged quantum vacuum. If we would employ a more realistic evolutionary model with no static \emph{in} or \emph{out} regions, an approximate definition of particles can be made by selecting those mode solutions of the field equation that come in some sense ``closest'' to Minkowski space limit. Physically this might be envisaged as a construction that `�`least disturbs'' the field by the expansion and in turn leads to the concept of ``adiabatic states" (introduced for the scalar fields long time ago \cite{PAR},
then put on rigorous mathematical footing \cite{LUD}
and later on extended to Dirac fields \cite{HOL}).

In future work, one could also cope with the degradation of the stored
information by intervening from time to time and actively correcting the
contents of the memory during the
evolution of the universe. In this direction, channel capacities taking into account this
possibility have been introduced in \cite{MRW}.
In another direction, and much more speculatively, one might attempt to find a meaningful notion
for entanglement-assisted communication in our physical scenario
by considering Einstein-Rosen bridges along the lines of \cite{MS13}
or entanglement between different universe's eras, related to dark energy \cite{CLM}.

\subsection*{Acknowledgments}
RP would like to thank Jonathan~P.~Dowling and the Hearne Institute for
Theoretical Physics, Louisiana State University, for the kind hospitality.
RP and SM are grateful to Shahpoor Moradi for helpful
discussions at the early stage of this work.
MMW acknowledges support from the Department of Physics and Astronomy
at Louisiana State University, from
the DARPA Quiness Program through US Army Research Office award W31P4Q-12-1-0019, and
from the NSF\ under Award No.~CCF-1350397.

\appendix

\section{Proof of Theorem~\ref{theo1}}
\label{sec:theorem1-proof}

The two dimensional trade-off region of Theorem \ref{theo1} is a special case of 
a theorem determining the triple trade-off region where in addition to $C$ and $Q$ also the net rate $E$ 
of entanglement consuption/generation is considered.

First we recall that the triple trade-off region for any quantum channel $\mathcal{A}%
_{A^{\prime}\rightarrow B}$\ is given by a union of polyhedra, each of which
is specified by the following formulas \cite{W13book,WH10b}:
\begin{eqnarray}
C+2Q  &  \leq I\left(  AX;B\right)  _{\rho},\label{eq:triple-1}\\
Q+E  &  \leq I\left(  A\rangle BX\right)  _{\rho},\\
C+Q+E  &  \leq I\left(  X;B\right)  _{\rho}+I\left(  A\rangle BX\right)_{\rho}.
\label{eq:triple-3}%
\end{eqnarray}

\begin{theorem}
\label{theo1gen} 
The single-letter triple trade-off region  (\ref{eq:triple-1})-(\ref{eq:triple-3}) for the qubit amplitude damping channel 
is the union of the following polyhedra over all $p_0, q_x,\nu_x\in\left[  0,1\right]$ for $x \in \{0,1\}$ and where 
$p_1 = 1 - p_0$ and
$p \equiv \sum_{x\in\{0,1\}} p_x q_x$:
\begin{eqnarray}
C+2Q &  \leq h_{2}\left(  \eta p\right)  +\sum_{x\in\{0,1\}}
p_x \, [{\sf g}\left(  q_x,1,\nu_x\right)  
-{\sf g}\left(q_x,1-\eta,\nu_x\right)]  ,\nonumber\\
Q+E &  \leq  \sum_{x\in\{0,1\}} p_x \, [{\sf g}\left(  q_x,\eta,\nu_x\right)  -{\sf g}\left(  q_x,1-\eta,\nu_x\right)]  ,
\nonumber\\
C+Q+E &  \leq  h_{2}\left(  \eta p\right)  -\sum_{x\in\{0,1\}} p_x  \, {\sf g}\left(  q_x,1-\eta,\nu_x\right) ,
\nonumber
\end{eqnarray}
where
\begin{equation*}
{\sf g}\left(  q,z,\nu\right)  \equiv  h_{2}\left(  \frac{1+\sqrt{\left(
1-2 z q\right)^{2}+4 z \nu^2 q (1-q)  }}{2}\right).
\end{equation*}
\end{theorem}

\textit{Proof.} From Refs.~\cite{W13book,WH10b} we have that the so-called
\textquotedblleft quantum dynamic capacity formula\textquotedblright%
\ characterizes the optimization task set out in (\ref{eq:triple-1}%
)-(\ref{eq:triple-3}) (i.e., the task of computing the boundary of the region
specified by (\ref{eq:triple-1})-(\ref{eq:triple-3})). That is, we should
optimize the quantum dynamic capacity formula for all non-negative values of
the Lagrange multipliers $\lambda$ and $\mu$ and doing so allows us to
simplify the form of ensembles necessary to consider in the computation of the
boundary of the region. The quantum dynamic capacity formula is given by%
\begin{eqnarray}
\max_{\left\{  p_{X}\left(  x\right)  ,\rho^{x}\right\}  }\left(  I\left(
AX;B\right)  _{\rho}+\lambda I\left(  A\rangle BX\right)  _{\rho}+\mu\left[
I\left(  X;B\right)  _{\rho}+I\left(  A\rangle BX\right)  _{\rho}\right]
\right),\nonumber\\\label{qdf}%
\end{eqnarray}
with the entropies referring to the state of (\ref{rhoXAB}). As detailed
in \cite{W13book,WH10b}, this is equivalent to
\begin{eqnarray}
I\left(  AX;B\right)  _{\rho}+\lambda I\left(  A\rangle BX\right)  _{\rho}%
+\mu\left[  I\left(  X;B\right)  _{\rho}+I\left(  A\rangle BX\right)  _{\rho
}\right]  \nonumber\\
=(1+\mu)H(B)_{\rho}+H(A|X)_{\rho}+\lambda H(B|X)_{\rho}-(1+\mu+\lambda
)H(E|X)_{\rho}, \label{qdf2}
\end{eqnarray}
where the various von Neumann entropies $H$ can be specified as follows.


A general input qubit density operator for the system $A^{\prime}$ has a
matrix representation as follows:
\begin{equation}
\rho^{x}=
\left(\begin{array}{cc}
\langle1|\rho^{x}|1\rangle & \langle1|\rho^{x}|0\rangle\\
\langle0|\rho^{x}|1\rangle & \langle0|\rho^{x}|0\rangle
\end{array}\right)
=
\left(\begin{array}{cc}
1-q_{x} & \gamma_{x}\\
\gamma_{x}^{\ast} & q_{x}%
\end{array}\right)
,\label{eq:qubit-dens-op}%
\end{equation}
where $q_{x}\in\lbrack0,1]$ and $\gamma_{x}\in\left[  0,\sqrt{q_{x}-q_{x}^{2}%
}\right]  $. Sending the qubit density operator (\ref{eq:qubit-dens-op})
through the amplitude damping channel $\mathcal{A}$ of (\ref{Kraus-A}) leads
to the following state at the output:
\begin{equation}
\mathcal{A}\left(  \rho^{x}\right)  =
\left(\begin{array}{cc}
1-\eta q_{x} & \sqrt{\eta}\gamma_{x}\\
\sqrt{\eta}\gamma_{x}^{\ast} & \eta q_{x}%
\end{array}\right).
\label{eq:qubit-dens-op-out}
\end{equation}
Then, referring to the state in (\ref{rhoXAB}), the output entropy $H(B)$ is
\[
H\left(  \sum_{x}p_{X}(x)\mathcal{A}(\rho^{x})\right)  ,
\]
while the conditional entropy $H\left(  A|X\right)  $ is
\begin{equation}
\sum_{x}p_{X}\left(  x\right)  h_{2}\left(  \frac{1+\sqrt{\left(
1-2q_{x}\right)  ^{2}+4\left\vert \gamma_{x}\right\vert ^{2}}}{2}\right)  .
\end{equation}
Furthermore, the conditional entropy $H\left(  B|X\right)  $ (of the output
given which state is input) is as follows:
\begin{equation}
\sum_{x}p_{X}\left(  x\right)  h_{2}\left(  \frac{1+\sqrt{\left(  1-2\eta
q_{x}\right)  ^{2}+4\eta\left\vert \gamma_{x}\right\vert ^{2}}}{2}\right)  .
\end{equation}
On the other hand, sending the qubit density operator in
(\ref{eq:qubit-dens-op}) through the channel $\tilde{\mathcal{A}}$
complementary to the amplitude damping channel $\mathcal{A}$ leads to the
following state at the environment:%
\begin{equation}
\tilde{\mathcal{A}}\left(  \rho^{x}\right)  =%
\left(\begin{array}{cc}
1-\left(  1-\eta\right)  q_{x} & \sqrt{1-\eta}\gamma_{x}\\
\sqrt{1-\eta}\gamma_{x}^{\ast} & \left(  1-\eta\right)  q_{x}%
\end{array}\right).
\end{equation}
Then, the conditional entropy $H\left(  E|X\right)  $ (of the environment
given which state is input) is
\begin{equation}
\sum_{x}p_{X}(x)h_{2}\left(  \frac{1+\sqrt{\left(  1-2\left(  1-\eta\right)
q_{x}\right)  ^{2}+4\left(  1-\eta\right)  \left\vert \gamma_{x}\right\vert
^{2}}}{2}\right)  .
\end{equation}
As discussed in Refs.~\cite{WH10b,W13book}, any simplification of the quantum
dynamic capacity formula can be helpful in reducing the space of parameters
over which we need to optimize. So our first aim is to simplify this formula
for the case of the amplitude damping channel. To this end we can always
augment an ensemble $\rho$ of the form in (\ref{rhoXAB}) to become
\begin{equation}
\sum_{x,j}\frac{1}{2}p_{X}\left(  x\right)  \left\vert x\right\rangle
\left\langle x\right\vert _{X}\otimes\left\vert j\right\rangle \left\langle
j\right\vert _{J}\otimes\left(  Z^{j}\rho^{x}Z^{j}\right)  _{A^{\prime}},
\end{equation}
where $Z$ is the Pauli $Z$ operator. This augmentation can only increase
communication rates due to the covariance of the amplitude damping channel
with respect to $\left\{  I,Z\right\}  $. Let $\sigma_{XJABE}$ denote the
corresponding classical-quantum state that results from purifying each state
in the $A^{\prime}$ system and then sending the $A^{\prime}$ system through an
isometric extension of the channel. That is,
\begin{eqnarray}
\sigma_{XJABE}\equiv\sum_{x,j}\frac{1}{2}p_{X}\left(  x\right)  \left\vert
x\right\rangle \left\langle x\right\vert _{X}\otimes\left\vert j\right\rangle
\left\langle j\right\vert _{J}\otimes\mathcal{U}_{A^{\prime}\rightarrow
BE}^{\mathcal{A}}\left(  Z^{j}|\phi^{x}\rangle_{AA^{\prime}}\langle\phi
^{x}|Z^{j}\right),\nonumber\\
\end{eqnarray}
with $\mathcal{U}_{A^{\prime}\rightarrow BE}^{\mathcal{N}}$ an isometric
extension of the channel $\mathcal{A}_{A^{\prime}\rightarrow B}$. We then have
an upper bound for the r.h.s.~of (\ref{qdf2}), namely
\begin{eqnarray}
&  \!\!\!\!\!\! (1+\mu)H(B)_{\rho}+H(A|X)_{\rho}+\lambda H(B|X)_{\rho}-(1+\mu
+\lambda)H(E|X)_{\rho}\nonumber\\
&  =(1+\mu)H(B)_{\rho}+H(A|XJ)_{\sigma}+\lambda H(B|XJ)_{\sigma}%
-(1+\mu+\lambda)H(E|XJ)_{\sigma}\nonumber\\
&  \leq
(1+\mu)h_{2}(\eta p)+H(A|XJ)_{\sigma}+\lambda H(B|XJ)_{\sigma}%
-(1+\mu+\lambda)H(E|XJ)_{\sigma}\nonumber\\
&  =(1+\mu)h_{2}(\eta p)+\sum_{x}p_{X}\left(  x\right)  \Bigg[h_{2}\left(
\frac{1+\sqrt{\left(  1-2q_{x}\right)  ^{2}+4\left\vert \gamma_{x}\right\vert
^{2}}}{2}\right)  \nonumber\\
&  \ \ \ \ \ \ \ \ \ \ +\lambda h_{2}\left(  \frac{1+\sqrt{\left(  1-2\eta
q_{x}\right)  ^{2}+4\eta\left\vert \gamma_{x}\right\vert ^{2}}}{2}\right)
\nonumber\\
&  \ \ \ \ \ \ \ \ \ \ -\left(  1+\mu+\lambda\right)  h_{2}\left(
\frac{1+\sqrt{\left(  1-2\left(  1-\eta\right)  q_{x}\right)  ^{2}+4\left(
1-\eta\right)  \left\vert \gamma_{x}\right\vert ^{2}}}{2}\right)
\Bigg],\label{1stineq}\nonumber\\
\end{eqnarray}
where the inequality follows from concavity of entropy and defining
\begin{equation}
p\equiv\sum_{x}p_{X}\left(  x\right)  q_{x}.
\end{equation}
Other steps follow from the covariance of the amplitude damping channel with
respect to $I$ and $Z$ operations.

As a consequence of (\ref{1stineq}), we see that to compute (\ref{qdf}), it
suffices to optimize the following function of $\left\{  \left(  p_{X}\left(
x\right)  ,q_{x},\gamma_{x}\right)  \right\}  $ for fixed values of $\lambda$
and $\mu$:
\begin{eqnarray}
&\left(  1+\mu\right)  h_{2}\left(  \eta p\right)  +\sum_{x}p_{X}\left(
x\right)  \Bigg[h_{2}\left(  \frac{1+\sqrt{\left(  1-2q_{x}\right)
^{2}+4\left\vert \gamma_{x}\right\vert ^{2}}}{2}\right)  \nonumber\\
&\hspace{2cm}+\lambda h_{2}\left(
\frac{1+\sqrt{\left(  1-2\eta q_{x}\right)  ^{2}+4\eta\left\vert \gamma
_{x}\right\vert ^{2}}}{2}\right) \nonumber \\
&\hspace{1cm}-\left(  1+\mu+\lambda\right)  h_{2}\left(  \frac{1+\sqrt{\left(
1-2\left(  1-\eta\right)  q_{x}\right)  ^{2}+4\left(  1-\eta\right)
\left\vert \gamma_{x}\right\vert ^{2}}}{2}\right)  \Bigg]. 
\nonumber\\
\label{eq:optimize-function}
\end{eqnarray}
Clearly, it suffices to take $\gamma_{x}$ real because the above function
depends only on the magnitude of $\gamma_{x}$.

First we argue that it is not necessary to consider distributions
$p_{X}\left(  x\right)  $ over more than two letters, and in order to do so,
we can apply the Fenchel-Eggelston-Carath\'{e}odory theorem often used in the
information theory literature for such purposes \cite{EK12}.
That is, we will show that to every probability distribution $p_{X}\left(
x\right)  $ over an arbitrary number of letters, there exists a probability
distribution $p_{X^{\prime}}\left(  x^{\prime}\right)  $ over just two letters
that achieves the same values of the function in (\ref{eq:optimize-function})
for fixed values of $\lambda$ and $\mu$.

Indeed, recall that the Fenchel-Eggelston-Carath\'{e}odory theorem states that
any point in the convex closure of a connected compact set in $S\subset
\mathbb{R}^{d}$ can be represented as a convex combination of at most $d$
points in $S$ (see e.g. \cite{EK12}). So, let us define the following two
functions of the parameters$~q$ and$~\gamma$:
\begin{eqnarray}
F_{0}\left(  q,\gamma\right)   &  \equiv q, \\
F_{1}(q,\gamma) &  \equiv h_{2}\left(  \frac{1+f_{0}(q,\gamma)}{2}\right)  +\lambda
h_{2}\left(  \frac{1+f_{1}(q,\gamma)}{2}\right) \nonumber\\ 
&-(1+\lambda+\mu)h_{2}\left(
\frac{1+f_{2}(q,\gamma)}{2}\right),
\label{eq:F1def}
\end{eqnarray}
with
\begin{eqnarray}
f_{0}(q,\gamma)\equiv &  \sqrt{(1-2q)^{2}+4\gamma^{2}},\\
f_{1}(q,\gamma)\equiv &  \sqrt{(1-2\eta q)^{2}+4\eta\gamma^{2}},\\
f_{2}(q,\gamma)\equiv &  \sqrt{(1-2(1-\eta)q)^{2}+4(1-\eta)\gamma^{2}}.
\end{eqnarray}
The functions $F_{0}$ and $F_{1}$ are continuous in $q$ and $\gamma$, and the
intervals $q\in\left[  0,1\right]  $ and $\gamma\in\left[  0,\sqrt{q-q^{2}%
}\right]  $ are connected and compact, so that the images of these functions
are connected and compact as well (the images taken together being in
$\mathbb{R}^{2}$). Thus, by applying the Fenchel-Eggelston-Carath\'{e}odory
theorem, we can conclude that there exists a probability distribution
$p_{X^{\prime}}\left(  x^{\prime}\right)  $ over just two letters such that
for $i\in\left\{  0,1\right\}  $%
\begin{equation}
\sum_{x}p_{X}\left(  x\right)  F_{i}\left(  q_{x},\gamma_{x}\right)
=\sum_{x^{\prime}}p_{X^{\prime}}\left(  x^{\prime}\right)  F_{i}\left(
q_{x^{\prime}},\gamma_{x^{\prime}}\right)  .
\end{equation}
Finally, the function of interest in (\ref{eq:optimize-function}) is a
continuous function of $\sum_{x^{\prime}}p_{X^{\prime}}\left(  x^{\prime
}\right)  F_{i}\left(  q_{x^{\prime}},\gamma_{x^{\prime}}\right)  $ for
$i\in\left\{  0,1\right\}  $ so that we can conclude that a probability
distribution on just two letters suffices for the optimization.

This concludes the proof of Theorem \ref{theo1gen} and in turn of Theorem \ref{theo1}.
$\hfill\blacksquare$

\bigskip

\section{Proof of Theorem~\ref{theo2}}
\label{sec:theorem2-proof}

Also Theorem \ref{theo2} can be seen as a special case of an analogous Theorem involving the triple
trade-off region.

\begin{theorem}
\label{theo2gen} 
The single-letter triple trade-off region  (\ref{eq:triple-1})-(\ref{eq:triple-3}) for the qubit amplitude damping channel 
when $\eta\geq 1/2$ is the union of the following polyhedra over all $p,\nu\in\left[  0,1\right]$:
\begin{eqnarray}
C+2Q &  \leq h_{2}\left(  \eta p\right)  +{\sf g}\left(  p,1,\nu\right)  -{\sf g}\left(
p,1-\eta,\nu\right)  ,\nonumber\\
Q+E &  \leq {\sf g}\left(  p,\eta,\nu\right)  -{\sf g}\left(  p,1-\eta,\nu\right)  ,\nonumber\\
C+Q+E &  \leq h_{2}\left(  \eta p\right)  -{\sf g}\left(  p,1-\eta,\nu\right) ,\nonumber
\end{eqnarray}
where ${\sf g}\left(  p,z,\nu\right)$ is defined in Theorem~\ref{theo1}
and it can be
achieved with the following ensemble 
\begin{eqnarray}
&\frac{1}{2}\left\vert 0\right\rangle \left\langle 0\right\vert _{X}\otimes%
\left(\begin{array}{cc}
1-p & \nu\sqrt{p\left(  1-p\right)  }\\
\nu\sqrt{p\left(  1-p\right)  } & p
\end{array}\right)
_{A^{\prime}}\nonumber\\
&+\frac{1}{2}\left\vert 1\right\rangle \left\langle 1\right\vert
_{X}\otimes
\left(\begin{array}{cc}
1-p & -\nu\sqrt{p\left(  1-p\right)  }\\
-\nu\sqrt{p\left(  1-p\right)  } & p
\end{array}\right)
_{A^{\prime}},\nonumber
\label{optimal}
\end{eqnarray}
with $p,\nu\in\left[  0,1\right]  $.
\end{theorem}

To prove Theorem~\ref{theo2gen} we have to show that ensembles of the following simplified form optimize
(\ref{eq:optimize-function}):
\begin{eqnarray}
&\frac{1}{2}\left\vert 0\right\rangle \left\langle 0\right\vert _{X}\otimes%
\left(\begin{array}{cc}
1-p & \nu\sqrt{p\left(  1-p\right)  }\\
\nu\sqrt{p\left(  1-p\right)  } & p
\end{array}\right)
_{A^{\prime}}\nonumber\\
&+\frac{1}{2}\left\vert 1\right\rangle \left\langle 1\right\vert
_{X}\otimes
\left(\begin{array}{cc}
1-p & -\nu\sqrt{p\left(  1-p\right)  }\\
-\nu\sqrt{p\left(  1-p\right)  } & p
\end{array}\right)
_{A^{\prime}}.\nonumber
\end{eqnarray}
This is equivalent to showing that for every $\{ p_X(x), q_x\}_{x\in\{0,1\}}$ such that
$\sum_x p_X(x) q_x = p$, there exists 
a value of $\nu$ such that
\begin{eqnarray}
&  \sum_{x}p_{X}\left(  x\right)  \Bigg[h_{2}\left(  \frac{1+\sqrt{\left(
1-2q_{x}\right)  ^{2}+4\left\vert \gamma_{x}\right\vert ^{2}}}{2}\right)
\nonumber\\
&+\lambda h_{2}\left(  \frac{1+\sqrt{\left(  1-2\eta q_{x}\right)  ^{2}%
+4\eta\left\vert \gamma_{x}\right\vert ^{2}}}{2}\right)
\nonumber\\
&-\left(  1+\mu+\lambda\right)   h_{2}\left(
\frac{1+\sqrt{\left(  1-2\left(  1-\eta\right)  q_{x}\right)  ^{2}+4\left(
1-\eta\right)  \left\vert \gamma_{x}\right\vert ^{2}}}{2}\right)\Bigg]\nonumber\\
&  \leq h_{2}\left(  \frac{1+\sqrt{\left(  1-2p\right)  ^{2}+4\nu^{2}p(1-p)}%
}{2}\right)  \nonumber\\
&+\lambda h_{2}\left(  \frac{1+\sqrt{\left(  1-2\eta p\right)
^{2}+4\eta\nu^{2}p(1-p)}}{2}\right)  \nonumber\\
&  -\left(  1+\mu+\lambda\right)  h_{2}\left(  \frac{1+\sqrt{\left(
1-2\left(  1-\eta\right)  p\right)  ^{2}+4\left(  1-\eta\right)  \nu
^{2}p(1-p)}}{2}\right).\label{eq:29}
\end{eqnarray}


Let us have a closer look at the function $F_{1}$ of Eq.(\ref{eq:F1def}). Its first derivative with
respect to $\gamma$ is as follows:
\begin{eqnarray}
\ln2\;\frac{\partial F_{1}(q,\gamma)}{\partial\gamma}&=
(1+\mu+\lambda)\frac{2(1-\eta)\gamma}{f_{2}}\ln\frac{1+f_{2}}{1-f_{2}}\nonumber\\
&-\lambda\frac{2\eta\gamma}{f_{1}}\ln\frac{1+f_{1}}{1-f_{1}}-\frac{2\gamma
}{f_{0}}\ln\frac{1+f_{0}}{1-f_{0}}\,.
\end{eqnarray}
This is a linear function of $\mu$, hence we can determine a critical value of
$\mu$ below (resp.~above) which $\frac{\partial F_{1}(q,\gamma)}%
{\partial\gamma}$ is always negative (resp.~positive). It is given by
\begin{equation}
\mu^{\ast}=-(1+\lambda)
+\frac{1}{\frac{1-\eta}{f_2}
\ln\frac{1+f_{2}}{1-f_{2}}}\Bigg[
\lambda\frac{\eta}{f_{1}}\ln\frac{1+f_{1}}{1-f_{1}}+\frac{1}{f_{0}}\ln\frac{1+f_{0}}{1-f_{0}}\Bigg]\,.
\end{equation}
The second derivative of $F_{1}$ with respect to $q$, in turn, is equal to
\begin{eqnarray}
&\ln 2\;\frac{\partial^{2} F_{1}(q,\gamma)}{\partial q^{2}}=
-\frac{4}{f_{0}^{2}}\,\left[  \frac{2\gamma^{2}}{f_{0}}\ln{\frac{1+f_{0}%
}{1-f_{0}}}+\frac{(1-2q)^{2}}{1-f_{0}^{2}}\right] \nonumber\\
&-4\,\lambda\,\frac{\eta^{2}}{f_{1}^{2}}\,\left[  \frac{2\eta\gamma^{2}}{f_{1}}%
\ln{\frac{1+f_{1}}{1-f_{1}}}+\frac{(1-2\eta q)^{2}}{1-f_{1}^{2}}\right]\nonumber\\
&+4\,(1+\mu+\lambda)\frac{(1-\eta)^{2}}{f_{2}^{2}}\left[  \frac{2(1-\eta)\gamma^{2}}{f_{2}%
}\ln{\frac{1+f_{2}}{1-f_{2}}}+\frac{(1-2(1-\eta)q)^{2}}{1-f_{2}^{2}}\right].\nonumber
\end{eqnarray}
This is also a linear function of $\mu$, and there exists a critical value of $\mu$
below (resp.~above) which $\frac{\partial^{2} F_{1}(q,\gamma)}{\partial q^{2}}$
is always negative (resp.~positive). It is given by
\begin{eqnarray}
\mu^{\ast\ast}=-(1+\lambda)&+\frac{\frac{f_{2}^{2}}{(1-\eta)^{2}}}
{\frac{{1-\eta}}{f_{2}}\ln\frac{1+f_{2}}{1-f_{2}}+\frac{f_2^2-4(1-\eta)\gamma^{2}}{2\gamma^{2}(1-f_{2}^{2})}}
\Bigg[\lambda\frac{\eta^{3}}{f_{1}^{3}}\ln\frac{1+f_{1}}{1-f_{1}}%
+\frac{1}{f_{0}^{3}}\ln\frac{1+f_{0}}{1-f_{0}}\nonumber\\
&+\lambda\frac{\eta^{2}}{2\gamma^{2}f_{1}^{2}}\frac{f_1^2-4\eta\gamma^{2}}{1-f_{1}^{2}}
+\frac{1}{2\gamma^{2}{f_0^2}}\frac{f_0^2-4\gamma^{2}}{(1-f_{0}^{2})}\Bigg].
\end{eqnarray}
By inspection, it follows that $\mu^{\ast}\leq\mu^{\ast\ast}$ for $\eta\ge\frac{1}{2}$ (for $\eta<\frac{1}{2}$ one can always find a large enough value of $\lambda$ that invalidate the condition). Anyway this is the only relevant regime for our purposes since for $\eta<\frac{1}{2}$ the quantum capacity of the amplitude damping channel vanishes. Let us then distinguish the
following two situations:


\subsection{$\mu\leq\mu^{\ast\ast}$, i.e. $F_{1}$ is concave with respect to
$q$}\label{i}

In this case $F_{1}$ is a monotonic function of $\gamma$. It is 
decreasing with increasing $\gamma$ for $\mu\leq\mu^{\ast}\leq\mu^{\ast\ast}$
and increasing with increasing $\gamma$  for
$\mu^{\ast}\leq\mu\leq\mu^{\ast\ast}$.

Nevertheless (remembering that it suffices to consider two letters) if we take
two arbitrary points $(q_{0},\gamma_{0})$ and $(q_{1},\gamma_{1})$ in the
$q,\gamma$ plane and suppose w.l.g. that $\gamma_{1}>\gamma_{0}$, we have
\begin{eqnarray}
F_{1}(q_{1},\gamma_{1})  &  \leq F_{1}(q_{1},\gamma_{0}),\qquad(  F_{1}
\;\mathrm{decreasing \; vs}\; \gamma),\nonumber\\
F_{1}(q_{1},\gamma_{1})  &  \ge F_{1}(q_{1},\gamma_{0}),\qquad( F_{1}
\;\mathrm{increasing \; vs}\; \gamma),\nonumber
\end{eqnarray}
hence
\begin{eqnarray}
&  p_{0} F_{1}(q_{0},\gamma_{0})+p_{1}  F_{1}(q_{1},\gamma_{1}) \leq p_{0}
F_{1}(q_{0},\gamma_{0})+p_{1}  F_{1}(q_{1},\gamma_{0}),\nonumber\\
&  (  F_{1} \;\mathrm{decreasing \; vs}\; \gamma),\nonumber\\
&  p_{0}  F_{1}(q_{0},\gamma_{0})+p_{1}  F_{1}(q_{1},\gamma_{1})\leq p_{0}
 F_{1}(q_{0},\gamma_{1})+p_{1} F_{1}(q_{1},\gamma_{1}),\nonumber\\
&  ( F_{1} \;\mathrm{increasing \; vs}\; \gamma).\nonumber
\end{eqnarray}
However, by the concavity of $F_{1}$ with respect to $q$ we can further write
\begin{eqnarray}
&  p_{0} F_{1}(q_{0},\gamma_{0})+p_{1}  F_{1}(q_{1},\gamma_{1})\leq  F_{1}%
(p_{0}q_{0}+p_{1}q_{1},\gamma_{0}),\\
&  ( F_{1} \;\mathrm{decreasing \; vs}\; \gamma),\nonumber\\
&  p_{0} F_{1}(q_{0},\gamma_{0})+p_{1} F_{1}(q_{1},\gamma_{1})\leq F_{1}%
(p_{0}q_{0}+p_{1}q_{1},\gamma_{1}),\\
&  ( F_{1} \;\mathrm{increasing \; vs}\; \gamma). \nonumber
\end{eqnarray}
So this proves (\ref{eq:29}) for this case. Notice that when $F_{1}$ is a decreasing
function of $\gamma$ the optimal value of $\gamma$ is $0$, while when $F_{1}$
is an increasing function of $\gamma$ the optimal value of $\gamma$ is the
maximum allowed one, i.e. $\sqrt{q-q^{2}}$.


\subsection{$\mu\ge\mu^{\ast\ast}$, i.e. $F_{1}$ is convex with respect to
$q$}\label{ii}

In this case $F_{1}$ is a monotonic increasing function of $\gamma$. Hence, we
should look for a suitable value of $\gamma$, say $\tilde{\gamma}$, such that
the following inequality (equivalent to (\ref{eq:29}))
\begin{equation}
\label{equiv29}p_{0} F_{1}(q_{0},\gamma_{0})+p_{1} F_{1}(q_{1},\gamma_{1})\leq
F_{1}(p_{0}q_{0}+p_{1}q_{1},\tilde{\gamma}), 
\end{equation}
is satisfied for any arbitrary points $(q_{0},\gamma_{0})$ and $(q_{1}%
,\gamma_{1})$ in the $q,\gamma$ plane (again remembering that it suffices to
consider two letters).

Since $F_{1}$ becomes increasingly convex with increasing $\mu$, the worst situation is
represented by the limit as $\mu\to\infty$, where
\begin{equation}
\label{Japprox}F_{1}(q,\gamma)\approx-\mu\, h_{2}\left(  \frac{1+f_{2}%
(q,\gamma)}{2}\right)  .
\end{equation}
To be on the safe side, let us consider $\gamma_{0}=\gamma_{1}=0$ where the
difference between the chord on the l.h.s.~of (\ref{equiv29}) and the function
$F_{1}(p_{0}q_{0}+p_{1}q_{1},0)$ is maximum. There, the worst situation is
represented by $q_{0}=0$, $q_{1}=1$ (hence $q=p_{1}$), for which we have
\begin{equation}
(1-q) F_{1}(0,0)+q F_{1}(1,0)\leq F_{1}(q,\tilde{\gamma}).
\end{equation}
Taking $\tilde{\gamma}=\sqrt{q-q^{2}}$ and accounting for (\ref{Japprox}) this
gives
\begin{equation}
q\mu\, h_{2}\left(  1-(1-\eta)q\right)  \ge0,
\end{equation}
which always holds true.

This concludes the proof of Theorem \ref{theo2gen} and in turn of Theorem \ref{theo2}.
$\hfill\blacksquare$

\bigskip

\begin{figure}[ptb]
\centering
\includegraphics[width=4.6in]{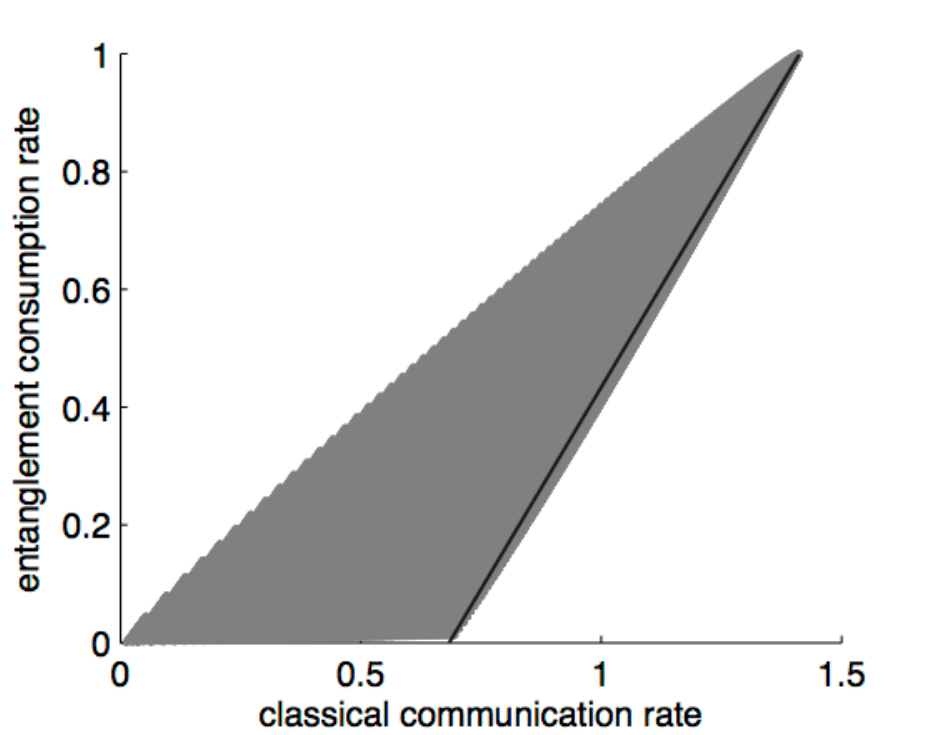}\caption{A comparison of a
trade-off coding strategy (blue points)\ versus a time-sharing strategy (red
line)\ for an amplitude damping channel with transmissivity $\eta=0.75$. The
figure demonstrates that an ensemble of the form in Theorem \ref{theo2gen}
outperforms a naive time-sharing strategy between the product-state classical
capacity and the entanglement-assisted classical capacity.}%
\label{fig:ce-tradeoff}%
\end{figure}

As consequence of Theorem \ref{theo2gen}, 
also the communication strategy involving entanglement  
results quite different from a naive time-sharing one and outperforms it 
(see Figure~\ref{fig:ce-tradeoff}).

\end{document}